\documentclass[review]{elsarticle}

\usepackage{lineno,hyperref}
\modulolinenumbers[5]

\journal{Journal of \LaTeX\ Templates}

\begin{document}

\begin{frontmatter}

\title{URu$_2$Si$_2$ under intense magnetic fields:\\ from hidden order to spin-density wave}

\author[address1]{W. Knafo\corref{mycorrespondingauthor}}
\cortext[mycorrespondingauthor]{Corresponding author}

\author[address2,address3]{D. Aoki}
\author[address1,address4]{G.W. Scheerer}
\author[address1]{F. Duc}
\author[address5]{F. Bourdarot}
\author[address6]{K. Kuwahara}
\author[address7]{H. Nojiri}
\author[address8]{L.-P. Regnault}
\author[address2]{J. Flouquet}

\address[address1]{Laboratoire National des Champs Magn\'{e}tiques Intenses, UPR 3228, CNRS-UPS-INSA-UGA, 143 Avenue de Rangueil, 31400 Toulouse, France}
\address[address2]{Service Photonique, Electronique et Ing\'{e}nierie Quantiques, Universit\'{e} Grenoble Alpes et Commissariat \`{a} l\'\ Energie Atomique, INAC, 17 rue des Martyrs, 38054 Grenoble, France}
\address[address3]{Institute for Materials Research, Tohoku University, Ibaraki 311-1313, Japan}
\address[address4]{DQMP, University of Geneva, 1211 Geneva 4, Switzerland}
\address[address5]{Service de Mod\'{e}lisation et d'Exploration des Mat\'{e}riaux, Universit\'{e} Grenoble Alpes et Commissariat \`{a} l\'\ Energie Atomique, INAC, 17 rue des Martyrs, 38054 Grenoble, France}
\address[address6]{Institute of Quantum Beam Science, Ibaraki University, Mito 310-8512, Japan}
\address[address7]{Institute for Materials Research, Tohoku University, Sendai 980-8578, Japan}
\address[address8]{Institut Laue-Langevin, 71 Avenue des Martyrs, CS 20156, 38042 Grenoble, France}

\begin{abstract}
A review of recent state-of-the-art pulsed field experiments performed on URu$_2$Si$_2$ under a magnetic field applied along its easy magnetic axis
$\mathbf{c}$ is given. Resistivity, magnetization, magnetic susceptibility, Shubnikov-de Haas, and neutron diffraction experiments are presented,
permitting to emphasize the relationship between Fermi surface reconstructions, the destruction of the hidden-order and the appearance of a
spin-density wave state in a high magnetic field.

\end{abstract}


\end{frontmatter}

Over the last four decades, the physics of URu$_2$Si$_2$ has revealed a unique richness in the family of strongly-correlated electrons systems \cite{mydosh11}.
In spite of a huge experimental and theoretical effort, none has been able to propose a consensual description of its low temperature ground state,
which develops below the transition temperature $T_0=17.5$~K, and whose order parameter remains unknown. In this hidden-order phase, strong intersite
magnetic fluctuations have been observed by inelastic neutron scattering at two particular wavevectors $\mathbf{k}_0=$~(0~0~1), which is equivalent to
(1~0~0), and $\mathbf{k}_1=$~(0.6~0~0) \cite{broholm91,bourdarot14}. A change of the carrier mobility was found to coincide
with the establishment of the hidden order \cite{LeRDawson89,Kasahara07,Bel04,Santander09}, being a first indication of
the interplay between the Fermi surface, the hidden-order and, thus, the magnetic properties of this itinerant magnet. Under hydrostatic pressure,
signatures of a phase transition at the critical pressure $p_c\simeq 0.5$~GPa have been observed by thermal expansion and neutron diffraction,
revealing the pressure-induced stabilization of long-range antiferromagnetic ordering of moments of amplitude $0.3-0.4$~$\mu_b/$U, with the
wavevector $\mathbf{k}_0$ \cite{Bourdarot05,Amitsuka07,Villaume08}. Shubnikov-de Haas quantum oscillations of the magnetoresistivity
indicated almost similar Fermi surfaces in the
low-pressure hidden-order and high-pressure antiferromagnetic phases, leading to the proposition that the hidden order, although of unknown
nature, has the same periodicity with wavevector $\mathbf{k}_0$ than the high-pressure antiferromagnetic order \cite{Hassinger10}. We note that the low-pressure
and low-field hidden-order phase has initially been labeled as an antiferromagnetic phase \cite{broholm87}, following the observation of a small magnetic moment
of amplitude $\simeq0.02-0.03$~$\mu_b/$U ordering with the wavevector $\mathbf{k}_0$ at temperatures below $T_0$. However, this small moment cannot
explain the large entropy associated with the transition and has been later assigned as non-intrinsic and due to sample inhomogeneity \cite{Niklowitz10}. When a
magnetic field $\mathbf{H}$ is applied along the easy magnetic axis $\mathbf{c}$ of this Ising system, a cascade of three first-order transitions
in a narrow magnetic field window was reported in early high-field experiments \cite{deVisser86,Sugiyama99}, indicating the destruction of the hidden-order phase in fields higher
than 35~T, the stabilization of field-induced phases in magnetic fields between 35 and 39~T, and the setting-up of a paramagnetic polarized regime
in fields higher than 39~T.

In this paper, we present a review of recent high-field experiments performed on high-quality URu$_2$Si$_2$ single crystals using state-of-the-art
pulsed magnetic field experiments at the LNCMI-Toulouse high-field facility \cite{Scheerer12,Scheerer14} and at the ILL-Grenoble
neutron source \cite{Knafo16}. Magnetization has been measured
by compensated coils and resistivity by the four-point technique using 60-T pulsed magnets at the Toulouse site. Neutron scattering has been carried
out using a transportable 40-T pulsed magnet on the triple axis spectrometer IN22 (CRG-CEA) at the ILL. For all magnetization and resistivity
measurements presented here, the magnetic field has been applied along the easy magnetic axis $\mathbf{c}$ of URu$_2$Si$_2$. For the neutron
scattering experiment, a field slightly tilted by 4.2~$^\circ$ from the $c$-axis was applied, with no incidence on the magnetic properties
expected for $\mathbf{H}\parallel\mathbf{c}$ \cite{Scheerer12b}.

\begin{figure}[b]
\includegraphics[scale=0.6]{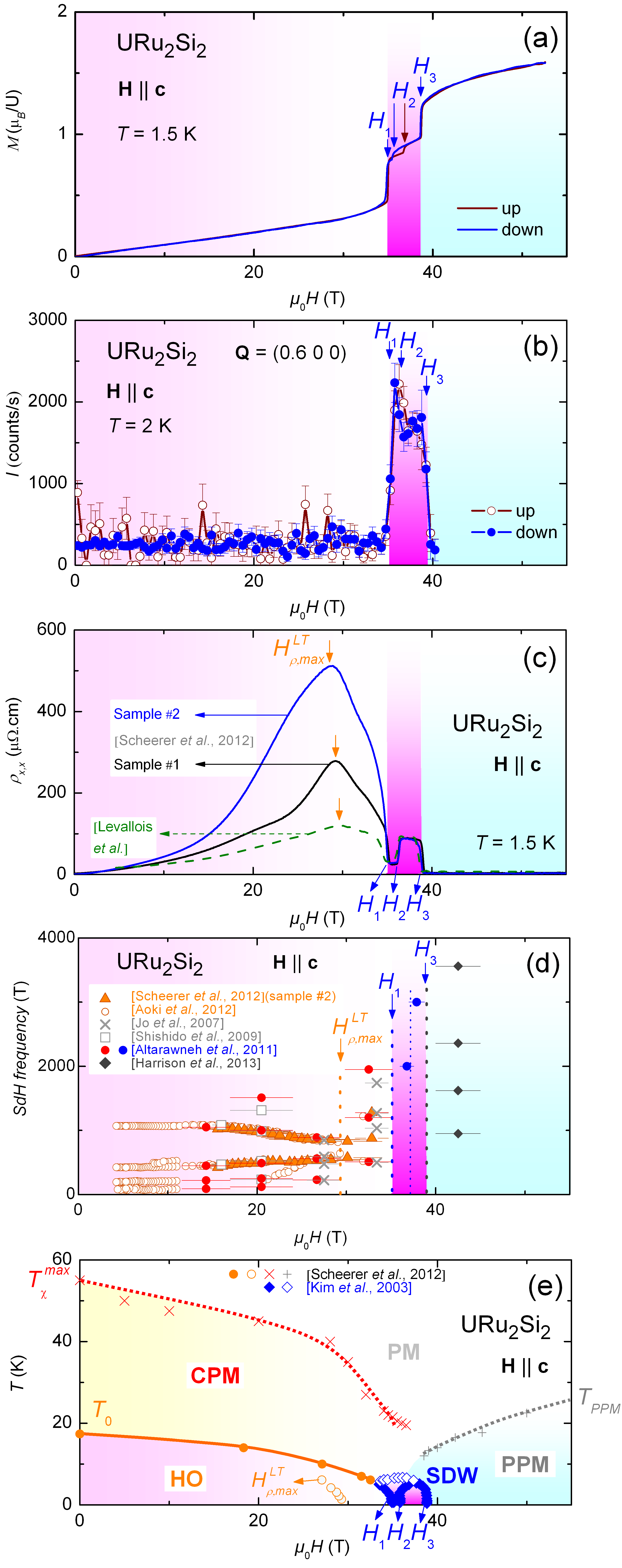}
\caption{(a) Magnetization at $T=1.5$~K (from Ref. \cite{Scheerer12}), (b) Neutron diffraction intensity at $\mathbf{Q}=(0.6 0 0)$ and $T=2$~K
(from Ref. \cite{Knafo16}),
(c) Resistivity of samples of different RRRs at $T=1.5$~K (from Refs. \cite{Scheerer12,Levallois09}), and (d) Shubnikov-de-Haas frequencies
(from Refs. \cite{Scheerer14,Aoki12,Jo07,Shishido09,Altarawneh11,Harrison14}), of URu$_2$Si$_2$
in a magnetic field $\mathbf{H}\parallel\mathbf{c}$. (e) Magnetic field - temperature phase diagram obtained from resistivity
and magnetization (from Refs. \cite{Scheerer12,Kim03}) measurements on URu$_2$Si$_2$ in a magnetic field $\mathbf{H}\parallel\mathbf{c}$.}
\label{fig1}
\end{figure}

\begin{figure}[b]
\includegraphics[scale=0.4]{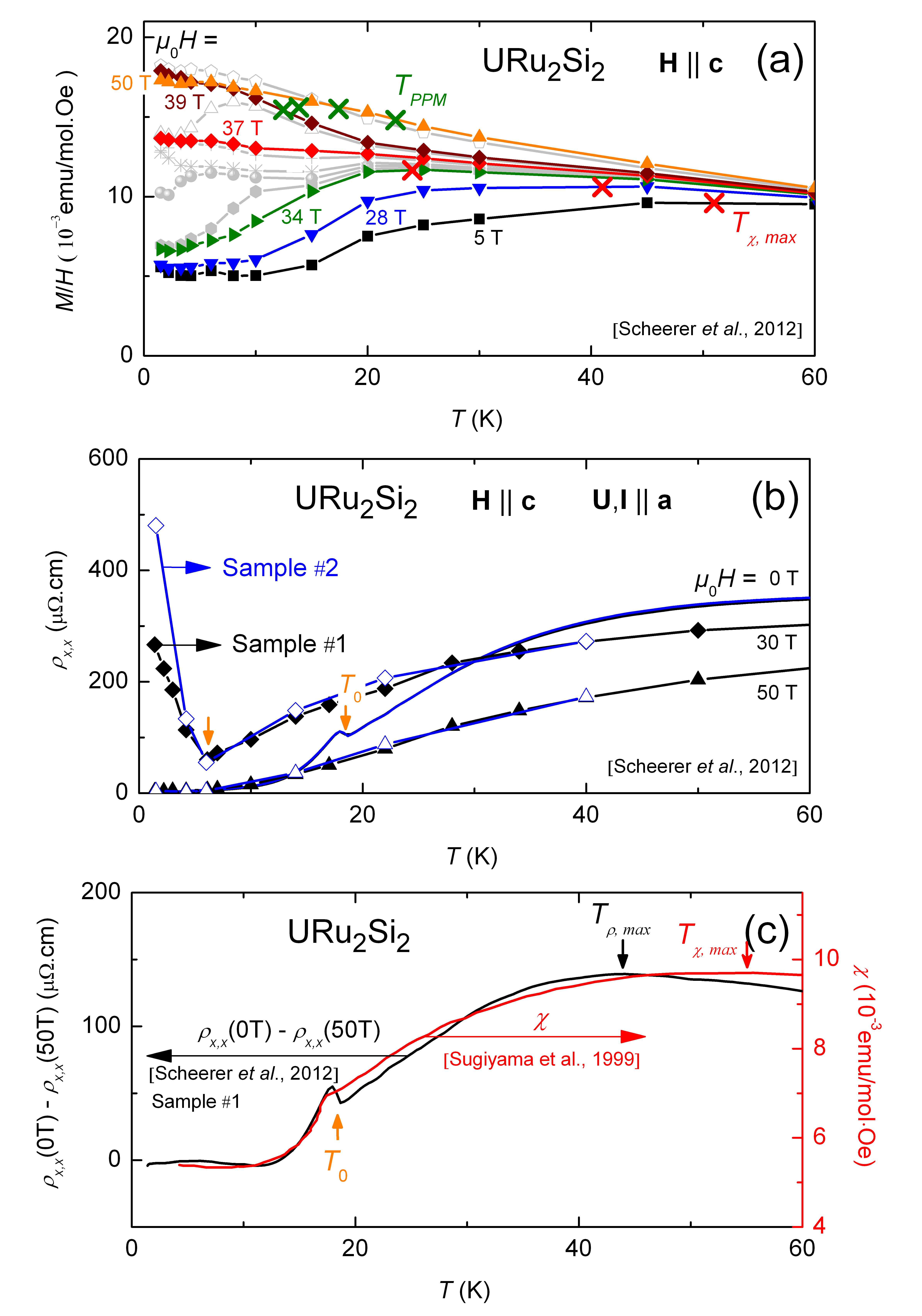}
\caption{(a)  Magnetization divided by the magnetic field $M/\mu_0H$ versus temperature at various magnetic fields
$\mathbf{H}\parallel \mathbf{c}$ (of 5, 20, 28, 32, 34, 34.5, 35, 35.5, 36, 36.5, 37, 37.5, 38, 28.5, 39, 40, 45, and 50~T)
(from Ref. \cite{Scheerer12}), (b) Resistivity $\rho_{x,x}$ versus temperature at $\mu_0H=0$, 30, and 50~T, for samples $\sharp1$ and $\sharp2$
(from Ref. \cite{Scheerer12}),
(c) Comparison of $\rho_{x,x}(T,\rm{0T})-\rho_{x,x}$$(T,\rm{50T})$ (from Ref. \cite{Scheerer12}) and the magnetic susceptibility $\chi(T)$
(from Sugiyama \textit{et al.} \cite{Sugiyama99}) versus temperature, of URu$_2$Si$_2$ in a magnetic field
$\mathbf{H}\parallel\mathbf{c}$.}
\label{fig2}
\end{figure}

Figure \ref{fig1}(a) presents the magnetization $M$ of URu$_2$Si$_2$ measured at $T=1.5$~K in magnetic fields $\mathbf{H}\parallel\mathbf{c}$. The
three first-order phase transitions are associated with step-like variations of the magnetization at the critical fields $\mu_0H_1=35$~T,
$\mu_0H_2=36/37$~T (rising/falling fields), and $\mu_0H_3=39$~T. In fields smaller than $H_1$, an almost linear increase of $M(H)$ is associated
with a large magnetic susceptibility ($\chi\simeq5\cdot10^{-3}$~emu/mol.Oe, see Figure \ref{fig2}(a)) typical of a heavy-fermion behavior. The
hidden-order is destroyed at $H_1$ and is replaced by field-induced phases for $H_1<H<H_3$, where a magnetization plateau corresponds to approximately
half of the total variation of $M$ between $H_1$ and $H_3$ (the step in the magnetization at $H_2$ is much smaller than those at
$H_1$ and $H_3$). For $H>H_3$, the magnetization reaches a large value $>1.3$~$\mu_B/$U characteristic of a polarized paramagnetic regime, and
continues to slowly increase with field, probably because of remaining unquenched magnetic fluctuations.

As shown in Figure \ref{fig1}(b), a neutron diffraction elastic Bragg peak at the wavevector $\mathbf{k}_1$ develops for $H_1<H<H_3$ (measurement
at the neutron momentum transfer $\mathbf{Q}=$~(0.6~0~0), which corresponds to the wavevector $\mathbf{k}_1$ via the relation
$\mathbf{Q}=\mathbf{\tau}+\mathbf{k}_1$, where $\mathbf{\tau}=$~(0~0~0) is a structural Bragg position). As detailed in Ref. \cite{Knafo16},
this Bragg peak
is the signature of a spin-density wave, i.e., a sine-modulation of magnetic moments, with an amplitude $2M(\mathbf{k}_1)\simeq0.5\pm0.05$~$\mu_B/$U
at $\mu_0H=36$~T. This amplitude is related with the variation $\Delta M\simeq0.4-0.5$~$\mu_B/$U in the magnetization between 36~T and just above
$H_3=39$~T, which is driven by the field-induced alignment (parallel to the field) of the moments ordered with wavevector $\mathbf{k}_1$ for $H<H_3$.
A small decrease of the
neutron intensity at $\mu_0H_2=36/37$~T, with a Hysteresis similar than in the magnetization measurement (see Fig. \ref{fig1}(a)),
indicates a subtle change at $H_2$ in the magnetic structure. Further high-resolution neutron diffraction experiments are needed to
determine the field-variation of the magnetic structure in the spin-density wave state.

The magnetoresistivity $\rho_{x,x}$ versus magnetic field of three different samples is presented in Fig. \ref{fig1}(c) (from Refs. \cite{Scheerer12,Levallois09}). The three transitions at
$H_1$, $H_2$, and $H_3$ are associated with sharp step-like variations delimiting two plateaus in the magnetoresistivity, where the value of
$\rho_{x,x}$ is smaller for $H_1<H<H_2$ than for $H_2<H<H_3$. While $\rho_{x,x}$ is almost sample-independent for $H>H_1$, its value is strongly
sample-dependent for $H<H_1$, i.e., in the hidden-order phase. As discussed in Refs. \cite{Scheerer12,Scheerer14}, the larger the residual resistivity ratio, the larger the
field-induced variation of $\rho_{x,x}$ is, indicating that the magnetoresistivity is mainly controlled by an orbital effect, the field-induced
cyclotron motion of the carriers, in the hidden order phase. This strong field-variation of $\rho_{x,x}$ is a consequence of the large carrier
mobility in the hidden-order state. A maximum in $\rho_{x,x}$ at the crossover field $H_{\rho,max}^{LT}\simeq30$~T (LT is used for "low temperature")
indicates a progressive reduction of the carrier mobility in the proximity of the transition field $H_1$ where the hidden order collapses. This
reduction of carrier mobility is related with a Fermi surface reconstruction, as discussed below.

Fig. \ref{fig1}(d) summarizes a large set of Fermi surface studies \cite{Scheerer14,Aoki12,Jo07,Shishido09,Altarawneh11,Harrison14} performed using Shubnikov-de Haas quantum oscillations of the resistivity in
URu$_2$Si$_2$ under high magnetic fields applied along $\mathbf{c}$. In this graph, the Shubnikov-de Haas frequencies extracted from Fourier
transforms of the quantum oscillations are plotted as a function of the magnetic field. In magnetic fields up to $\simeq15$~T, the frequencies
$F_\eta\simeq90$~T, $F_\gamma\simeq200$~T, $F_\beta\simeq400$~T, $F_\alpha\simeq1100$~T associated with the Fermi surface bands $\eta$, $\gamma$,
$\beta$, and $\alpha$, respectively, are almost unchanged. Above 15~T, a cascade of field-induced changes in the Fermi surface is observed.
In the hidden-order phase, a progressive variation of the frequencies is induced in a large crossover regime going from 15 to 30~T, and is followed by a
more sudden change of the frequencies at $\simeq30$~T, which coincides with the field $H_{\rho,max}^{LT}$ at the maximum of the orbital
magnetoresistivity. Fermi surface reconstructions have also been observed in thermoelectric measurements \cite{pourret13}.
A cascade of Fermi surface changes inside the hidden-order phase has been evidenced by Shubnikov - de Haas oscillations
by different groups. However, due to the difficulty to perform Fourier transforms in limited field ranges, the different analyzes led to slightly
different spectra \cite{Scheerer14,Aoki12,Jo07,Shishido09,Harrison14}. From studies performed in steady fields up to 45 T by Altarawneh
\textit{et al.} \cite{Altarawneh11} and Harrison \textit{et al.} \cite{Harrison14}, Shubnikov-de Haas frequencies
characteristic of the Fermi surface in the spin-density wave state and in the polarized paramagnetic regime have also
been extracted, showing that the transitions at $H_1$, H$_2$, and $H_3$ are accompanied by Fermi surface reconstructions. As seen in Figure
\ref{fig1}(d), the field-induced cascade of Fermi surface crossovers and reconstructions is accompanied by a general trend: an increase of the
Shubnikov-de Haas frequencies and, thus, of the associated Fermi surface volumes.

In Figure \ref{fig2}(a), the magnetic susceptibility $\chi=M/H$ is plotted as function of temperature for different field values. The temperature
$T_{\chi}^{max}$ defined at the maximum of $\chi(T)$ for $\mu_0H<35$~T delimits a low-field correlated paramagnetic regime. The temperature $T_{PPM}$
defined at the inflexion point of $\chi(T)$ for $\mu_0H>35$~T is the borderline of the high-field polarized paramagnetic regime. In Figure
\ref{fig2}(b), the resistivity $\rho_{x,x}$ is plotted as a function of temperature for two samples (samples $\sharp1$ and $\sharp2$) at the
magnetic field values $\mu_0H=0$, 30, and 50~T. The resistivity at zero field is mainly sample-independent (a small sample dependence is observed
at low temperatures, reflecting the different residual resistivity ratios, but cannot be seen in this graph). In agreement with the conclusions of
the low-temperature $\rho_{x,x}$ versus $H$ plot (Figure \ref{fig1}(c)), the $\rho_{x,x}$ versus $T$ plot in Figure \ref{fig2}(b) confirms that a
sample-dependent magnetoresistivity is observed only in the hidden-order phase, i.e., at temperatures below $T_0\simeq6$~K for $\mu_0H=30$~T. In a
magnetic field of 50~T, the electron-electron contribution to the low-temperature electrical resistivity is strongly reduced in comparison with that
at zero-field. We estimate by $\rho_{x,x}({\rm{50T}},T)$ the phononic contribution to the zero-field resistivity $\rho_{x,x}({\rm{0T}},T)$ of
URu$_2$Si$_2$, and by $\rho_{x,x}({\rm{0T}},T)-\rho_{x,x}({\rm{50T}},T)$ the purely electronic contribution to the zero-field resistivity. As shown in
Figure \ref{fig2}(c), the $T$-dependences of  $\rho_{x,x}({\rm{0T}})-\rho_{x,x}({\rm{50T}})$ and the magnetic
susceptibility $\chi$ are very similar, showing broad maxima at $\simeq45$ and 55~K, respectively. This indicates that both quantities in this
temperature range are controlled by the progressive setting of the correlated paramagnetic regime.

Figure \ref{fig1}(e) presents the magnetic field - temperature phase diagram of URu$_2$Si$_2$ in a field $\mathbf{H}\parallel\mathbf{c}$, constructed
from magnetization and magnetoresistivity data (from Refs. \cite{Scheerer12,Kim03}). Inside the hidden-order phase, the crossover field
$H_{\rho,max}^{LT}$ decreases with increasing temperature in a similar
manner than the hidden-order borderline, indicating that $H_{\rho,max}^{LT}$ is an intrinsic property of the hidden-order phase. At temperatures higher
than $T_0$, a broad maximum at the temperature $T_{\chi}^{max}$ in the magnetic susceptibility marks the onset of a
low-temperature heavy-Fermi-liquid plateau in the magnetic susceptibility, as in usual heavy-fermion paramagnets. Remarkably, in a large number of
heavy-fermion paramagnets (including URu$_2$Si$_2$) a scaling between $T_{\chi}^{max}$ and the field-induced pseudo-metamagnetic field $H_m$,
indicates that a single energy scale controls the correlated paramagnetic regime \cite{Aoki13}. The specificities of URu$_2$Si$_2$ are the appearance of its
hidden-order state below the temperature $T_0$ and of a field-induced spin-density wave beyond the
hidden-order phase. Under a magnetic field, both $T_0$ and $T_{\chi}^{max}$ vanish in the critical field area [35-39T] where the spin-density wave is
stabilized and above which a polarized paramagnetic regime is established. In the future, further efforts are needed to describe
the field-induced phases labelled II, III and V \cite{Kim03}. In particular, the question whether phase II (which develops below 6~K in the field window 34-38~T)
is a real phase of a crossover regime has been recently raised \cite{mydosh_pc}.

The interplay between the magnetism and the Fermi surface is a key to understand the electronic properties of URu$_2$Si$_2$. The
simultaneous field-induced changes of the magnetic and Fermi surface properties reported here are a direct illustration of this interplay.
Further, the hidden-order
state is characterized by strong magnetic fluctuations at the wavevectors $\mathbf{k}_0$ and $\mathbf{k}_1$, which have been identified as nesting
(or quasi-nesting) vectors of the Fermi surface \cite{Ikeda12}. When an external parameter is tuned, as pressure, uniaxial stress, Rh-doping, or magnetic field, long-range
magnetic order can be stabilized either with wavevector $\mathbf{k}_0$ (via pressure \cite{Bourdarot05,Amitsuka07,Villaume08}, uniaxial stress \cite{Kambe13,Bourdarot11}, Rh-doping
\cite{Yokoyama04,Oh07,Kim04}) or with the wavevector $\mathbf{k}_1$ (via a magnetic field $\mathbf{H}\parallel\mathbf{c}$ \cite{Knafo16}).
Within an itinerant picture of magnetism, one can speculate that the application of these tuning parameters leads to
modifications of the Fermi surface nestings, inducing the stabilization of long-range ordering with the wavevectors $\mathbf{k}_0$ or $\mathbf{k}_1$.
However, the fact that almost no Fermi surface change has been reported experimentally under pressure, while a cascade of
Fermi surface reconstructions were reported in a field $\mathbf{H}\parallel\mathbf{c}$, illustrates how subtle the properties of URu$_2$Si$_2$ are.
By revealing the relationship between the Fermi surface and magnetism of URu$_2$Si$_2$ under pressure and magnetic field, new
generations of band structure calculations (see also Refs. \cite{Denlinger01,Elgazzar09,Oppeneer10,Ikeda12,Suzuki14}) will surely
help describing quantitatively its properties and
perhaps solving the hidden-order problem. Another challenge would be to understand why an 'up-up-down'
ferrimagnetic structure is stabilized in high magnetic field in Rh-doped U(Ru$_{0.96}$Rh$_{0.04}$)$_2$Si$_2$ with the wavevector
$\mathbf{k}_2=$(2/3 0 0) \cite{Kuwahara13}, $\mathbf{k}_2$ being very close to the wavevector $\mathbf{k}_1$ of the field-induced
spin-density wave in pure URu$_2$Si$_2$.

This work was supported by Programme Investissements d\'\ Avenir under the program ANR-11-IDEX-0002-02, reference ANR-10-LABX-0037-NEXT.
Part of this work was funded by the ANR grant Magfins N$^{\circ}$ ANR-10-0431. H.N. acknowledges KAKENHI 23224009. D.A. acknowledges KAKENHI,
15H05745, 15H05882, 15H05884, 15K21732, 25247055. H.N. and D.A. acknowledge support by ICC-IMR.

\end{document}